# SRAMM: Short Read Alignment Mapping Metrics

Alvin Chon[1, 2] and Xiaoqiu Huang[1, 2]

[1]Bioinformatics and Computational Biology Graduate Program, Iowa State University, Ames, IA 50010, USA
[2]Department of Computer Science, Iowa State University, Ames, IA 50010, USA

*ABSTRACT*

*Short Read Alignment Mapping Metrics (SRAMM): is an efficient and versatile command line tool providing additional short read mapping metrics, filtering, and graphs. Short read aligners report MAPing Quality (MAPQ), but these methods generally are neither standardized nor well described in literature or software manuals. Additionally, third party mapping quality programs are typically computationally intensive or designed for specific applications. SRAMM efficiently generates multiple different concept-based mapping scores to provide for an informative post alignment examination and filtering process of aligned short reads for various downstream applications.*

*SRAMM is compatible with Python 2.6+ and Python 3.6+ on all operating systems. It works with any short read aligner that generates SAM/BAM/CRAM file outputs and reports 'AS' tags. It is freely available under the MIT license at http://github.com/achon/sramm.*

## 1. Introduction

Short read alignment is vital to almost all Bioinformatics pipelines and currently only MAPping Quality (MAPQ) [1] scores are universally reported in short read alignment, SAM/BAM, files [2]. However, MAPQ is a non-robust singular score representing the probability of the read or read pair's mapping alignment against a reference. Also, mapping quality as defined by MAPQ is distinctly a different measure than alignment score or probability of an alignment. Additionally, the genomic reference is not standardized across different applications leading to different alignments which may not be reflected in the score depending upon the alignment tool. Lastly, MAPQ suffers from differing implementations across various common tools such as BWA-MEM, Bowtie2, MOSAIK, VG [3, 4, 5, 6], etc and in general is not standardized. Therefore, there is a need for additional mapping metrics for downstream applications. Third party approaches to mapping quality work with read aligners [7], are re-implementations of read aligners [8], or are stand alone [9, 10]. Many have complex models or specific assumptions are not applicable to all pipelines [11, 12]. Lastly, many are computationally intensive and require as much time as the read aligners themselves. In contrast, SRAMM generates additional concept-based mapping scores: Mapped Identity (MI), Unique-Repeat (UR), and UnMapped (UM) scores to provide additional information that a user can translate into criteria for read filters. The mapping scores are based upon read alignment scores and not MAPQ. These scores allow for versatile filtering based upon different mapping concepts at the read alignment stage of analysis. The classic and common process is selecting uniquely mapping reads. We can filter reads exactly using a combination of adjustable MIs, URs, and number of alignments while MAPQ cannot be translated unambiguously and is read aligner dependent. Additionally, depending upon the experiment and downstream analysis, the concept of mapping can differ. In many cases, the MAPQ metric is neither suitable nor sufficient by itself as discussed in various Bioinformatic communities. Here we describe SRAMM, an efficient and versatile command line Python tool to address this need.

## 2. METHODS

### 2.1. Implementation

SRAMM is a lightweight command line executable to facilitate the ease of use on any system and in any pipeline scheme. It has three primary functions: generate metrics, filter results, and compute graphs. Currently it supports SAM/BAM/CRAM files through the pysam package [13]. The program is designed to perform any combination of the processes from command line arguments with optional parameters. The output is a tab delimited stats file containing read names, raw alignment scores, number of alignments, MAPQ, and the computed statistics. Additionally, graphs are generated using the matplotlib package [14] for the score distributions of the unfiltered or filtered data set. These outputs allow for visualization and examination of the short read data set. Lastly, additional filtering can be integrated into pipelines using samtools [14] or directly with command line tools on the output data files.

### 2.2. Metrics

SRAMM introduces 3 mapping scores: Mapped Identity (MI), Unique-Repeat (UR), and UnMapped (UM) in addition to retaining MAPQ and Number of Alignments. SRAMM accepts either single and paired end short reads therefore all metrics below have both single and paired end implementations. Here, we will generalize the equations and refer to the paired end version in text. The introduced mapping metrics are not designed to be used in a standalone manner due to their inherent ambiguity similar to MAPQ. They are intended to be used in combinations to allow for different mapping concepts to intersect providing unambiguous filtering or read selection.

In each run of this program, all reads should fall under the same alignment score distribution. They are not required to be the same length, but they should have the same alignment scoring distribution. By allowing reads with different scoring distributions, it makes human interpretation of equivalent scores ambiguous. In general, the read set must have the same alignment score range and scheme regardless of read length. Currently AS is typically reported as an integer, therefore this is not the general case. For example, reads of lengths 50 and 100 will typically have different AS ranges. But, if the aligner calculates AS in such a way that they are mapped to the same score distribution, then it is possible to combine reads. Generally, this is not true, so a trivial solution to this is to segregate reads of different lengths, run SRAMM on each set of reads creating normalized mapping score ranges, and then pool the results. Additionally, this approach can be used for different experimental conditions such as different read lengths, library preparation, sequencing applications, and other experimental variables.

Notation below is for paired end reads. The single end case is simpler as it would be the singular alignment score and not the sum of the pair of alignments and thus not explicitly shown.

Let $m_i$ denote the number of alignments of the read pair $i$ to the reference.

let $s_j$ denote the sum of the two scores for the alignment pair $j$, $1 <= j <= m_i$.

Let $S$ be the maximum of the $s_j$ values for all $i$ and $j$.

The mapping scores are defined from the above notation.

### 2.2.1. Mapped Identity

Mapped Identity score: The maximum alignment score for a given read pair, $i$, divided by the maximum possible alignment score of the data set, $S$. Range = [0, 1]. See equation 1 below.

$$MI_i = \frac{max(s_i^1, s_i^2, \ldots, s_i^{m_i})}{S}$$

This is the percent identity for the highest alignment pair. Conceptually, it is how well did the pair of reads map anywhere to the provided reference. This metric is not sensitive to how many alignments nor where they are in the reference, hence Mapped Identity. By itself, this mapping score is not very informative. But, when combined with MAPQ or UR, it can be very helpful in adding further selection criteria and removing ambiguity.

### 2.2.2. Unique-Repeat

Unique Repeat Score: The unique-ness to repeat-ness of a pair of reads, $i$, as a query against the reference. That is, a higher score denotes a unique mapping location whereas a low score means the read maps to multiple locations with varying degrees. Range = (0, 1]. This score is highly sensitive to the number of alignments or alignment pairs. See equation 2 below.

$$UR_i = \frac{max(s_i^1, s_i^2, \ldots, s_i^{m_i})}{Sm_i} + \frac{\sum_{j=1}^{m_i} S - s_i^j}{Sm_i}$$

If there is only one pair of alignments, then it is as unique as it is score divided by the max score which is MI. By the Bioinformatics common definition of a repeat, we have a lower score for more alignments. Additionally, each alignment of the score is weighted by their scores since a low alignment score means less confidence in the alignment itself, reducing the weight of that alignment as a possible repeat.

Given the equation, it should be noted that a value of 1 is a theoretical max and is only achieved when the pair of reads only maps to one location with perfect alignment or has only 1 alignment pair with scores of 0. This occurs frequently as aligners have score thresholds for reporting alignments which clip low score alignments and must be taken into consideration when running the short read aligner. Additionally, the above formula considers all alignments output by the aligner. Some aligners will output more possible alignment pairs than others which can lead to very different score distributions.

It should be noted that this score cannot be negative, however the closer the score is to 0 denotes a repeat in regards to the distribution of UR scores. If the read aligns to multiple places perfectly, then the score would be 1 divided by the number of alignments as more perfect alignments means more repeats which reduces the score. Most reads greater than 100bp will not map to multiple locations all with high scores unless there is some common sequence or repeated element. The distribution of this score in an experiment is highly dependent upon read length, scoring schema especially how it reports alternate alignments and the cutoff, and the reference sequences themselves. It is difficult to make cross experiment comparisons if many of the variables are different. However, given a pipeline with repeated samples, direct comparison of the distributions is meaningful.

Another key point is that UR by itself is ambiguous. The ambiguity is that the same score can come from read pairs that have a different number of alignments. A read pair with 2 alignments has an UR range of [0.5, 1.0) whereas a read pair with 3 alignments has an UR of [0.333, 1.0). Therefore, an arbitrary score of say 0.90 can come from a read pair with 2, 3, or any number of alignments. Conceptually, this is intended. In the example above, both are almost unique even though there are multiple alignments. That is, even though it has 2 or 3 alignments, the alternates are of such low alignment that they do not carry weight to the concept of being a repeat. However, that doesn't help a user who wants to purely filter reads by number of alignments. To further the selection criteria, we include the sheer number of alignment pairs. That way a user can select a range of alignments or a specific number and combine that with UR to get reads of interest.

### 2.2.3. UnMapped

UnMapped Score: A measure of how unmapped the read pair $i$ is. This is both based upon the number of alignments and the actual scores of the alignments, but heavily weighted by the scores. Range = [0,1]. See equation 3 below.

$$UM_i = 1 - \frac{\sum_{j=1}^{m_i} s_i^j}{Sm_i}$$

An unmapped pair of reads conceptually should have none too few alignments and they should all be of low score. This greatly depends on the aligner as most have a reporting threshold for alignments. Additionally, a pair of reads with perfect alignment scores would then have a score of 0 even if there are multiple alignments. In general, as there are more alignments with low scores, the UM score increases.

If there are few alignments, this distribution will appear similar to the left-right flipped MI distribution. As the proportion of reads with multiple alignments in the data set increases, the MI and UM distributions will diverge. This can be used as a general measure to see how many read pairs have multiple alignments.

Ideally this metric is used in coordination with the read aligner's alignment reporting parameters. Many read aligners have reporting thresholds and or handle multiple alignments differently, thus this metric expands that continuum. In general, we do not want a hard cutoff for reporting alignments that possibly affects reads of interest in a dataset.

## 3. RESULTS

SRAMM can process the alignments of 1M and 10M paired end reads at 101bp per read in less than 1 and 7 minutes respectively while using less than 1GB RAM on an Intel(R) Core(TM) i7-7500U CPU @ 2.70GHz using 2 threads. This includes stat generation, optional filtering, and optional graph generation. Due to I/O operations, namely reading the millions of alignments and writing statistics for each read, increasing the number of cores does not linearly improve performance even with increased memory usage. However, the number of reads stored in memory can be increased to reduce the number of I/O batch operations while increasing memory usage, but this process is still dependent upon system cache settings.

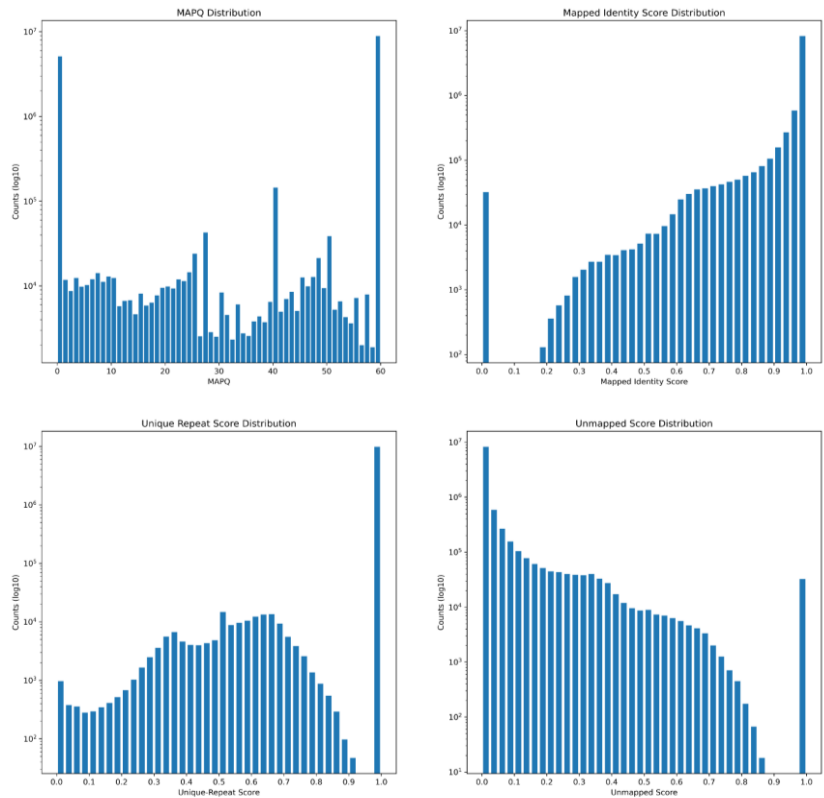

Figure 1. SRAMM metrics on 10M reads from NA12878

Graphs of SRAMM score distributions. 10M 101bp paired end reads were aligned using BWA-MEM and then processed with SRAMM. From top left to bottom right: MAPQ, MIs, URs, UMs.

### 3.1. Mapped Identity and Uniquely Mapped Reads

We present results for running SRAMM on 10M 101bp paired end reads from Illumina aligned to the hg38 [15] using BWA-MEM. The data is from NA12878 [16] and shows the general distribution of MAPQ and the metrics over the 10M reads. It is to be noted that many read pairs have multiple alignment pairs and BWA-MEM reported them.

We perform using samtools a MAPQ filter of greater than or equal to 60 and only one primary alignment. For BWA-MEM, MAPQ=60 is the maximum reported value. This filter step reduced the read count from 10M to 8,817,232 or 8.8M reads. It is common for a read pair to have multiple alignment pairs and still report a MAPQ of 60 due to the alternate alignments being low scoring. 8,870,228 read pairs had at least one alignment pair with a MAPQ of 60 which means approximately 53,000 read pairs had multiple alignments yet still had a MAPQ of 60. We then ran SRAMM on the resulting reads as seen in Figure 2.

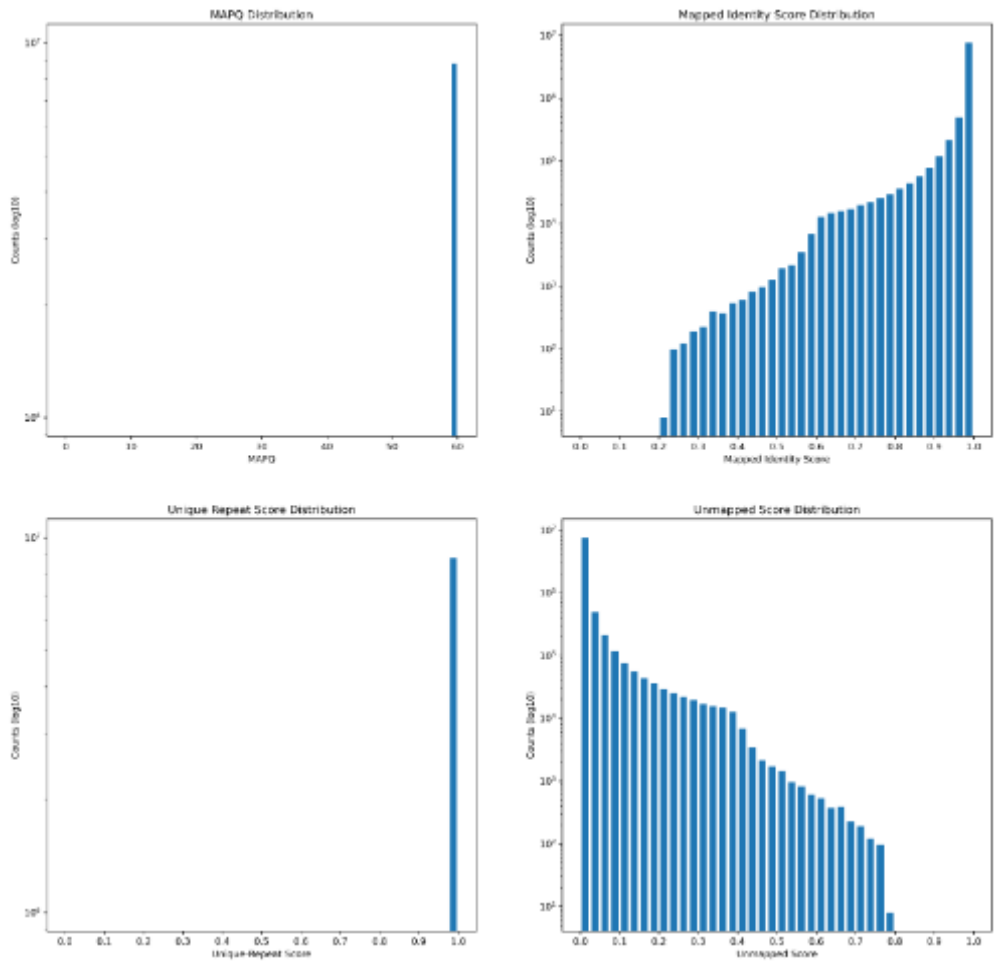

Figure 2. SRAMM metrics on NA12878 MAPQ=60, 1 alignment

Graphs of SRAMM score distributions. 8.8M 101bp paired end reads were aligned using BWA-MEM, filtered using samtools for MAPQ=60, and then processed with SRAMM. From top left to bottom right: MAPQ, MIs, URs, UMs.

From this figure it is evident that we filtered for MAPQ=60 and 1 alignment. The Unique-Repeat distribution is made up of solely values of 1.0 due to there being only 1 primary alignment. However, what is unusual is that the Mapped Identity distribution varies greatly. From the graph, most alignments average less than 0.90 or less than 90% sequence identity. Some reads even have such a % identity as low as 0.20 yet still are reported as MAPQ=60. This is due to the nature of the MAPQ calculation which is heavily dependent upon the quality score of the base in the read. While it may be a mismatch to the reference, the quality score is low enough that it does not impact MAPQ. Therefore, out of a 101bp read, even if 20 to 30 bps match, the mismatches are of such low quality that it can still generate a MAPQ of 60. Therefore, reads with % identities ranging from 0.20 to 1.0 all can report a MAPQ which may not be desired. An experiment may require high % identity which clearly may not be filtered for by using MAPQ.

## 4. CONCLUSIONS

We developed SRAMM and demonstrated its ability to quickly generate additional mapping statistics and related functionality. To date, there are numerous approaches to mapping quality that typically require customization for each application or overhead. Alternatively, SRAMM works on read aligner independent output for data examination, visualization, and mapping quality filtering of reads. A user can examine an experiment's distributions for various purposes such as comparing to previous experiments to validate that they are similar, or any number of filtering operations based upon user driven criteria. For example, using re-sequencing data, we can filter out all non-truly uniquely mapping reads by using a combination of MI and UR scores which is not feasible with MAPQ alone. This allows a user to select reads that map with a chosen identity percentage and only a primary alignment. Conversely, if we are interested in reads that did not align well or suspected repeat reads, we utilize a combination of low MI, low UR, and high UM scores along with the number of alignments. This combination of multiple mapping metrics is more versatile and informative than using MAPQ alone without being a computationally expensive addition to a pipeline. Therefore, our read aligner independent tool can be incorporated into any short read pipeline for quality control or filtering of reads for downstream applications.